\documentstyle[12pt]{article}
\def\ni{\noindent}

\newcommand{\mk}{|{\bf k}|}
\begin{document}

\begin{tabbing}
\`SUNY-NTG-93-10\\
\`January 1993
\end{tabbing}
\vbox to  0.8in{}

\centerline{\Large \bf Real photons from nonequilibrium QGP}
\vskip 2.5cm
\centerline{\large A.  Makhlin }
\vskip .3cm
\centerline{Department of Physics}
\centerline{State University of New York at
Stony Brook}
\centerline{Stony Brook, New York 11794}
\vskip 0.35in
\centerline{\bf Abstract}

We calculate the rate of the emission of the photons from the QGP. We base on
the real-time kinetic approach [1] without
an explicit assumption about a complete thermal
equilibrium in the emitting system.

\vfil
\noindent
${^\dagger}$ E-mail address: "makhlin@sbnuc"
\eject

\pagebreak

{\bf 1.} The subject of this note is as follows. In recent papers devoted
to  the rate of emission of the real photons
from the quark-gluon plasma one may  find  very different final answers.
 Most of them agree as for the main logarithmic terms which
appropriately estimate rate of emission only at unphysically large energies.
 Next terms
which come to be significant at the typical value of $E/T\sim 10$ different
authors write in manifestly different way [2,3]. It looks extremely surprising
even the different functional dependence from the basic parameters,
though all authors start from the same initial formulas.
I have seen no papers were the procedure of analytic approximation would
have been discussed in detail except paper [4] were expression for the
rate of emission was given in the form of one-dimensional integrals which
allow the reliable further analysis by very simple means.

{\bf 2.} Expression for the rate of emission,
 more accurate then the main logarithm,
is needed in connection with some modern scenario of the QGP formation which
predict  suppression of the quark
component against background of the hot glue
rather than complete thermalization of the QGP [5].

 For sake of completeness
 we shall start with reminding of the definitions and the main framework
 of calculations. For photon spectrum the rate of emission is given by [1,4]
 \begin{equation}
k^{0}{{dN_{\gamma}}\over{d{\bf k}d^{4}x}} =
   {{ig_{\mu\nu}}\over{2(2\pi)^{3}}} \pi^{\mu\nu}_{10}(-k),
 \end{equation}
with the usual definition of two currents correlation function
\begin{equation}
 \pi^{\mu\nu}_{10}(x,y) =
        i\langle T^{+}(j^{\mu}(x)S^{+}) T(j^{\mu}(x)S)\rangle,  \\
\end{equation}

 In what follows we use the same way of  derivation of the two loop expansion
of $\pi^{\mu\nu}_{10} $ as in [1,4] and denote
$\;\;g_{\mu\nu} \pi^{\mu\nu}_{10}(-k)\;\; $ as $\;\;\pi(k)$.

{\bf 3.} As the threshold of the first Born`s term and the
one-loop radiative corrections is $k^{2}>4m^{2}$ they are
 absent for real photons.
Physically, this mean that the two fermion states with the  time-like
momenta can not produce the massless photon. In the vertex which emits the
photon one of the fermion legs should carry the time-like momenta and
another - a space-like one.  The fermion field excitations with space-like
momentum can not belong to normal set of the normalizable  states and
therefore to contribute the density matrix.  They can be only virtual. So
only the real processes of the next perturbation order can contribute the
rate of emission. The desired virtuality of the fermion line in the
electromagnetic vertex comes from the intermediate state of the
annihilation or the Compton process.

It would have been inconsistent to include
 the Born's term with  the fermion lines
dressed by the one-loop self-energy in the whole scheme where the two-loop
and the next graphs are considered perturbatively. It will lead to an
over-count in the subset of graphs of the self-energy type and to the
under-count in the group of graphs of the vertex type.

In fact, an account for some elements of a collective   behavior in the
ensemble of the partons is needed only in a very narrow region of a
collinear geometry of the real processes which leads to the logarithmic
divergence of the emission rate. A simple restriction of the size of the
domain of  the coherence of the interaction between the partons reflects
the physics of the partonic ensemble and resolve the problem of a
divergence.

  This allows us to consider only the real processes, annihilation and
Compton, where the necessary virtuality is provided by the gluons.
In this case the trace of the electromagnetic
 polarization tensor  reads as [6]
 \begin{eqnarray}
\pi_{real}=-{ie^{2}g^{2}N_{c}C_{F} \over  2\pi^{5} } \int d^{4}pd^{4}q
\delta(q^{2})\delta[(p-k)^{2}-m^{2}]\delta[(p+q)^{2}]
 (SW_{a}+SW_{c})  \times \nonumber\\
\times \{1+ {A \over (p^{2}-m^{2})^{2}}+{B \over p^{2}-m^{2}}  +
{ C \over p^{2}-m^{2}+2kq}   \}
\end{eqnarray}
where  we denoted:
\begin{equation}
A=4m^{2},\;\;
B=3m^{2}+2(kq)-{ 2m^{4} \over (kq)},\;\;
C=m^{2}+{ 2m^{4} \over (kq)} .
\end{equation}

The expression in the curly  brackets is a sum of the
 squared moduli of the matrix
elements of the annihilation process, $q\bar{q}\rightarrow g\gamma$, or
Compton process, $qg\rightarrow q\gamma$ and
$\bar{q}g\rightarrow \bar{q}\gamma$. Just for the purpose of the following
calculations it is written not it terms of habitual Mandelstam variables
$(s,t,u)$. Specification of the process as well as their kinematic
boundaries are due to statistical weights.

In the  formal approximation when distributions of
initial particles are taken to be Boltzmann
ones the statistical weight of the annihilation process looks as
\begin{eqnarray}
SW_{a}=\theta(k_{0}-p_{0}) \theta(q_{0}+p_{0}) \theta(q_{0})
n_{F}(k_{0}-p_{0})n_{F}(q_{0}+p_{0})n_{B}(q_{0}) \approx \nonumber \\
\approx \theta(k_{0}-p_{0}) \theta(q_{0}+p_{0}) \theta(q_{0})
{e^{-(ku)/T}  \over e^{-(qu)/T}-1 }.
\end{eqnarray}
For the Compton rates the statistical weight is equal to
\begin{eqnarray}
SW_{c}=-\theta(p_{0}+q_{0}) \theta(p_{0}-k_{0}) \theta(-q_{0})
n_{F}(p_{0}+q_{0})[1-n_{F}(p_{0}-k_{0})]n_{B}(-q_{0})-   \nonumber \\
-\theta(-p_{0}-q_{0}) \theta(k_{0}-p_{0}) \theta(-q_{0})
n_{F}(k_{0}-p_{0})[1-n_{F}(-p_{0}-q_{0})]n_{B}(-q_{0})- \approx \nonumber \\
\approx -e^{-(ku)/T}[\theta(p_{0}+q_{0}) \theta(p_{0}-k_{0}) \theta(-q_{0})
[\exp((pu-ku)/T)+1]^{-1} - \nonumber \\
-\theta(-p_{0}-q_{0}) \theta(k_{0}-p_{0}) \theta(-q_{0})
[\exp((qu)/T)+1]^{-1}   ]
\end{eqnarray}
This approximation was proved to be quantitatively effective for the case
of the true thermal equilibrium of the emitting plasma.

We may also assume that quarks and gluons
are not in thermal equilibrium. Computer simulation of the parton
cascade usually produces something similar to
the Boltzmann distributions
damaged by the specific parameters $\zeta_{Q}$ and $\zeta_{G}$.
We adopt for quarks and gluons
\begin{equation}
n_{F}(p)=\zeta_{Q}e^{-pu/T},\;\;\;\;\;\;n_{B}(p)=\zeta_{G}e^{-pu/T},
\end{equation}
where $u^\mu$, the 4-velocity of the nonequilibrium partonic media,
fugacities $\zeta$ and temperature $T$ are very smooth functions of
space-time coordinates. Scenario of the hot glue leads to
$\zeta_{Q}<\zeta_{G}<1$  and all these quantities have no habitual
thermodynamical meaning.

In this parametrization of the partons distributions
which we shall name the partons cascade,
the statistical weight of the annihilation process with
emission of a gluon looks as
\begin{equation}
SW_{a}^{casc}=\approx \theta(k_{0}-p_{0}) \theta(q_{0}+p_{0}) \theta(q_{0})
\zeta_{Q}^{2} e^{-(ku)/T}  (e^{-(qu)/T}+\zeta_{G}e^{-2(qu)/T})
\end{equation}

For the Compton rate of the photon emission the statistical weight equals to
\begin{eqnarray}
SW_{c}^{casc}
\approx  -\zeta_{Q} \zeta_{G} \{ \theta(p_{0}+q_{0}) \theta(p_{0}-k_{0})
\theta(-q_{0}) e^{-pu/T}[1-\zeta_{Q} e^{-(pu-ku)/T}] + \nonumber \\
+\theta(-p_{0}-q_{0}) \theta(k_{0}-p_{0}) \theta(-q_{0})
e^{-ku/T} e^{(qu+pu)/T}[1-\zeta_{Q} e^{(qu+pu)/T}] \}
\end{eqnarray}
For any kind of these distributions we can perform an
exact integration over $p$ using the Breit reference system
where ${\bf k + q}=0$, c.m.s. of the reaction $q\bar{q}\rightarrow g\gamma$.

For the equilibrium distributions of the partons this integration
leads to
\begin{equation}
\pi_{ann}=-{ie^{2}g^{2}N_{c}C_{F} \over  4\pi^{4} } e^{-ku/T}
\int d^{4}q \delta(q^{2}) \theta(q_{0}) \theta[(k+q)^{2}-4m^{2}]
{ {\cal F}_{a}(kq) \over e^{qu/T}-1}
\end{equation}

 where

\begin{equation}
{\cal F}_{a}(x)= (1+ {2m^{2} \over x} - {2m^{4}
\over x^{2} }) \ln{ 1- \sqrt{1-2m^{2}/x}
\over  1+ \sqrt{1-2m^{2}/x} }
+(1+ {2m^{2} \over x}) \sqrt{1-{2m^{2} \over x} }
\end{equation}

For the Compton process it is reasonable to
start with a chain of changes of variables, $p\rightarrow -p+k-q$,
in the first term and $q \rightarrow -q-p$ in both terms.
After that we may also easily
perform the  integration over $p$  using the same Breit reference system.
It gives
\begin{equation}
\pi_{compt}=-{ie^{2}g^{2}N_{c}C_{F} \over  4\pi^{4} } e^{-ku/T}
\int d^{4}q \delta(q^{2}-m^{2}) \theta(q_{0}) { {\cal F}_{c}(kq) \over
e^{qu/T}+1}
\end{equation}
where $q$ is the momentum of the (anti)quark in the final state and
\begin{equation}
{\cal F}_{c}(x)=(1-{2m^{2}\over x}-{2m^{4}\over x^{2}})
\ln{m^{2}+2x \over m^{2}}
+{4m^{2} \over x }+{2x(m^{2}+x) \over (m^{2}+2x)^{2} }
\end{equation}

 In course of these    calculations we could easily learn that the energy $p_0$
of a virtual quark in the annihilation process equals to zero in
the Breit system. In the Compton process it equals to the ratio
$m^2 /2(k_0 + q_0 ) $ which is small for the hard photon also.  At least
in both cases the momenta of virtual quark  are space-like. Because of the high
energy of the photon  the 4-momentum $p^{\mu}$ of a virtual quark
is very close to the light cone in the rest frame of the media.

{\bf 4.} Further integration is somewhat tricky. Indeed,
for the photons $\pi_{compt}$ depends
only on one time-like 4-vector $u$, $u^{2}=1$.
So to continue calculations in a reasonable way we are to chose reference frame
$\vec{u}=0$. In this frame the integral of Eq.(9) takes form,
\begin{equation}
-\pi T \int_{0}^{\infty} d[\ln(1+e^{ -\sqrt{q^{2}+m^{2}} } )] q   \nonumber \\
\int^{1}_{-1} dz
{\cal F}_{c}(k_{0} \sqrt{q^{2}+m^{2}}+\mk q z)
\end{equation}
where  $k_{0}=\mk$ stands for $ku$.  The trick which allows one to reduce this
double integral to  a simple quadrature is as follows [4]:

i)change the variable $z$ for the new one, $\mk qz=\eta$;

ii) integration with respect to $q$ by parts;

iii) successive changes  of variables: $q=m\sinh \phi$, then $me^{\phi}/T=2y$.

The result reads as
\begin{eqnarray}
 \pi_{compt}=-{ie^{2}g^{2}N_{c}C_{F} \over  2\pi^{3} } e^{-ku/T} T^{2}
 \int_{0}^{\infty} dy \ln(1+e^{-y-{\lambda^{2} \over 4y}})\times \\
\times [(1-{\xi \over y}-
 {\xi^{2} \over 2y^{2}} )\ln(1+{4y \over \xi}) +{2\xi \over y}+
 {4y(\xi+2y) \over (\xi +4y)^{2} }  ]    \nonumber
\end{eqnarray}
where $\xi=m^2 /(ku)T \;$, $\lambda =m/T$ and singular
 point $y=0$ corresponds to the
infinite rapidity $\phi$ of the massless quark.

For the $q\bar{q}$ annihilation into gluon and real
photon, $k^{2}=0$, the angular integration is also performed in the rest
frame of the media and may be reduced to
\begin{equation}
\pi T \int_{0}^{\infty} d[\ln(1-e^{-q/T})] q \int^{1}_{-1}dz
\theta(q(k_{0}+\mk z)-2m^{2}){\cal F}_{a}((k_{0}+\mk z)q)
\end{equation}
After changing variable $z$ for $\eta=q(k_{0}+\mk z)$  integration over $q$
by parts  and changing $q$ for $y=(m^{2}/ku)q$ we arrive at
\begin{eqnarray}
\pi_{ann}=-{ie^{2}g^{2}N_{c}C_{F} \over  2\pi^{3} } T^{2} e^{-ku/T}
\int_{\xi}^{\infty} dy \ln(1-e^{-y}) \times \nonumber \\
\times [(1+{\xi\over y}-{\xi^{2} \over 2y^{2}})
 2{\rm cosh}^{-1}\sqrt{{y \over \xi }}+
(1+{\xi \over y})\sqrt{1-{\xi\over y}}  ]
 \end{eqnarray}

 It is easy to see that the spectrum of gluons in the annihilation process is
effectively cut from the above at gluon momenta $\sim  T$. So from a simple
kinematics it follows that at high photon energy, $E>>T$, the energies
of the initial quarks should be large, time-like and slightly above the
light cone.

The Compton process can be more precisely identified as the bremsstrahlung
of a photon due to the hard scattering of a quark on a gluon. Maximum rate
of the emission is produced by the infinite rapidity of the massless
quark in the final state.

{\bf 5.} Both annihilation and Compton rates as they are
given above  logarithmically
diverge   when quark mass is going to zero and the geometry of collision is
collinear.  In this case all the participants of the reaction  interact
infinitely long. This can not be true in the media and all the momenta should
be
cut  off from the below.

  The external cut-off should come from the smallest of the three lengths.
  The first one is the Compton wave-length $l_{c}\sim 1/m$. The second one is
  the mean free path defined via the ordinary or transport cross-section,
  $l_{mfp}\sim 1/g^2 T$ (or $\sim 1/g^4 T$). The third one is connected with
  the amplitude of the forward scattering, $l_{fs}\sim 1/gT \sim 1/m_{therm}$.
  As long as we keep quarks massless and consider coupling as the small
  parameter we must chose the length $l_{fs}$ as the smallest one.    So
  whenever we meet the collinear singularity the momenta of both real and
  virtual quarks should be cut off from the below at $\sim gT$.

  For the  production of photons with the energy $E$ much exceeding any of
  these scales the particular choice is not so significant within the
  reasonable accuracy. The cut-off mass appears only under the logarithm
  and its variation is not so valuable.

{\bf 6.} Having reached our
 goal to present the expression for the rates of emission from the
equilibrium plasma
 in the form of simple quadratures we can easily find them numerically.
 We can analyze the reliability of different analytic approximations and
compare
 them with those known in the literature.

 The way of calculations chosen  in the papers [2] and [3]
 had  followed paper [7] where at the very initial stage it
 was performed change of variables from original momenta to the invariant
 variables $(s,t,u)$. This trick produces  a set of delta-functions
 which are  naturally used to get rid of angular integrals.  This results
 in a cumbersome   four-dimensional integral which can be simplified only
 by assuming that $s>>t$. This assumption (as it was used in [7]) do not touch
 the main logarithmic term which is contributed basically by the low-angle
 scattering (small $t$).  But it violates the constant term which is mainly due
 to large-angle scattering, not to tell about the next terms. Though these are
 parametrically small, they are not small numerically in the physically actual
 region.

   I failed to trace out the way of analytic approximation in Ref.[3]
 but the final result  disagree even with the well established functional
 form of the leading logarithmic terms.

 Let us start with the annihilation mechanism of the real photons production.
 Notice that rate of photon emission depend upon a single parameter
 $\xi=m^{2}/(ku)T$ and contains the  logarithmic (collinear) divergence when $m
 \rightarrow 0$. Assuming the regularization by means of the mass cut-off
 at $m \sim gT$ we see that $\xi <<1$ at least because of $(ku)>>m \sim T$.

 My statement is that the approximation which picks up only the main logarithm
 of the small parameter $\xi$ fails at reasonable values of this parameter
(actually, $\xi \sim 0.1$). I will show that terms like $\xi \ln\xi$,
 $\xi \ln^{2}\xi$,... appear as well and that at reasonable values of $\xi$
 they essentially change the rate of emission.

 Let us consider the leading integral of $\pi_{ann}$,
\begin{equation}
-\int_{\xi}^{\infty} dy \ln(1-e^{-y})\ln{{4y \over \xi }}
\end{equation}
Its main part which is easily obtained by taking zero for the low limit
 equals to
\begin{equation}
{\pi^{2} \over 6} ( \ln{4 \over \xi} +1) +\{\int_{0}^{\infty} {y \ln y \;dy
\over e^{y}-1 } \approx -0.24\} \end{equation}
while estimation of the residue at small $\xi$ adds a term like
\begin{equation}
{1 \over 2}(\xi \ln{1 \over \xi} - \xi) \end{equation}
Estimation of the next terms is somewhat more complicated but it is very easy
to
understand the type of corrections: as the integrand has logarithm in the
nominator so the extra powers of $y$ in the denominator will produce the powers
of $\ln \xi$.

Just similar analysis can be performed for the Compton rate of emission.
The integral (15) is effectively cut from the below by
$y_{min} \sim \lambda^{2}/4 $ . The main term may be approximated as
\begin{equation}
\int_{\lambda^{2}/4}^{\infty}\ln(e^{-y}+1) (\ln{4y \over \xi} +{1 \over 2})
\end{equation}
Proceeding as above we get  the leading part,
\begin{equation}
{\pi^{2} \over 12}   (\ln{4 \over \xi} +{1 \over 2}    )  +
\{\int_{0}^{\infty}\ln(1+e^{-y})\ln y dy = -0.372\}
\end{equation}
and the corrections,
\begin{equation}
-{\lambda^{2} \over 4}(\ln {\lambda^{2}  \over 4}-1)\ln 2 +
{\pi^{2} \over 12} {3\lambda^{2} \over 8 } \ln 2
\end{equation}

The "reasonably small" value of the parameter $\lambda = m/T$ is ${\cal O} (1)$
and  the corrections are comparable with the main term    at the previous
value of $\xi \sim 0.1$.
The rest terms in  the  integral (15) will produce corrections with the
next powers of $\ln \lambda$, just as in the case of the annihilation .

This kind of analysis helps us to understand why it may be dangerous to
use the limit of the high photon energy in the many-dimensional integrals.
In this form it is hard to figure out the true structure of corrections.

At this circumstances it seems more reliable to use the quadratures
(15) and (17) as  the definitions without further approximations.
The result of numerical calculations is plotted
 at Fig.1 apart from the common
constant factor and the Boltzmann factor $\exp(-E/T)$. The difference
between exact function and the approximate ones is easily seen to be
almost constant at the photon energies exceeding 2 Gev. An approximation
given in Ref.[2] approaches the exact result only at $E\sim$100Gev. At
$E<$4Gev this approximation leads even to the wrong relation between
the annihilation and Compton rates.

{\bf 7.} The partons distributions (7) of the nonequilibrium regime lead  to
another expressions for the rate of emission. All technique of calculations
remains the same and the final results immediately follow from the next very
simple observation. Inspecting statistical weights given by Eqs. (5)
and (8)  we see that the following changes are required in the
annihilation channel:
\begin{equation}
[e^{qu/T}-1]^{-1} \rightarrow  e^{-qu/T} \rightarrow
\zeta_{Q}^{2}[e^{-qu/T} +\zeta_{G}e^{-2qu/T}]
\end{equation}
The middle of this string corresponds to the straightforward Boltzmann
approximation which results in the replacement of the  $\ln(1-e^{-y})$ in
Eq.(17)
by the $\;\; -e^{-y}$. Now the prescription is evident,
\begin{eqnarray}
\pi^{casc}_{ann}={ie^{2}g^{2}N_{c}C_{F} \over  2\pi^{3} } T^{2} e^{-ku/T}
\zeta_{Q}^{2} \int_{\xi}^{\infty} dy
[e^{-y} + {\zeta_{G} \over 2} e^{-2y} ]\times  \nonumber \\
\times [(1+{\xi\over y}-{\xi^{2} \over 2y^{2}})
 2\cosh^{-1}\sqrt{{y \over \xi }}+(1+{\xi \over y})\sqrt{1-{\xi\over y}}  ]
 \end{eqnarray}
The first exponent in the integrand corresponds to spontaneous emission of a
gluon, the second exponent with the extra factor $\zeta_{G}$ corresponds to the
 induced emission.

  The rate of photons production  in the Compton channel modifies in a
 similar way. A string of changes in the Eq.(12),
\begin{equation}
[e^{qu/T}+1]^{-1} \rightarrow  e^{-qu/T} \rightarrow
\zeta_{Q}\zeta_{G}[e^{-qu/T} -\zeta_{Q}e^{-2qu/T}]
\end{equation}
results in
\begin{eqnarray}
 \pi^{casc}_{compt}=-{ie^{2}g^{2}N_{c}C_{F} \over  2\pi^{3} } e^{-ku/T} T^{2}
 \zeta_{Q}\zeta_{G} \int_{0}^{\infty} dy
 [e^{-(y+{\lambda^{2} \over 4y})}-
 {\zeta_{Q} \over 2} e^{-2(y+{\lambda^{2} \over 4y})} ] \times \nonumber \\
 \times [(1-{\xi \over y}-
 {\xi^{2} \over 2y^{2}} )\ln(1+{4y \over \xi}) +{2\xi \over y}+
 {4y(\xi+2y) \over (\xi +4y)^{2} }  ]
\end{eqnarray}
 where the second exponent is due to the Pauli suppression of a quark in the
 final state.

The actual values of the fugacities which are the measure of the chemical
equilibrium are about $\zeta_Q \sim 0.5$ and $\zeta_G \sim 0.75$ [5]. In
order to find out the influence of the kinetic nonequilibrium produced by
the Boltzmann distributions we plot the rates given by (25) and (27) using
$\zeta_Q = \zeta_G =1 $ ( also apart from the common Boltzmann factor
$\exp(-E/T)$).

We consciously do not integrate the rates over the history of the system
which can be easily done for a simple form of the hydrodynamic background.
The very approach of nonequilibrium field kinetics [1,6] was designed in
order to use the nonequilibrium partons distributions generated in a
"realistic" cascade.

\vskip 1.5cm
I am indebted to G. Brown, E. Shuryak and the Nuclear Theory group at
SUNY at Stony Brook for continuous support.

 I am grateful to  E. Shuryak and I.Zahed for
many fruitful and helpful discussions.
\vskip 1.5cm

 \centerline{\bf REFERENCES}
\bigskip

 \ni [1] A.Makhlin, Preprint SUNY-NTG-92-11, Stony Brook, 1992  \\
 \ni [2] J.Kapusta, P. Lichard, D. Seibert: Phys. Rev. D44 (1991) 2774.\\
 \ni [3] R. Baier et al. Z.Phys.-C 53 (1992) 433. \\
 \ni [4] A. Makhlin: JETP 49 (1989) 238.\\
 \ni [5] E. Shuryak: Phys. Rev. Lett. 68 (1992) 3270. \\
 \ni [6] A.Makhlin, Preprint SUNY-NTG-93-20, Stony Brook, 1993  \\
 \ni [7] G. Staadt, W. Greiner, J. Rafelski: Phys.Rev. D33(1986)66.\\

\end{document}